# Ethical Guidelines for the Construction of Digital Nudges

Christian Meske
Freie Univeristät Berlin
christian.meske@fu-berlin.de

Ireti Amojo
Freie Univeristät Berlin
ireti.amojo@fu-berlin.de

## Abstract

*Under certain circumstances, humans tend to behave in irrational ways, leading to situations in which they make undesirable choices. The concept of digital nudging addresses these limitations of bounded rationality by establishing a libertarian paternalist alternative to nudge users in virtual environments towards their own preferential choices. Thereby, choice architectures are designed to address biases and heuristics involved in cognitive thinking. As research on digital nudging has become increasingly popular in the Information Systems community, an increasing necessity for ethical guidelines has emerged around this concept to safeguard its legitimization in distinction to e.g. persuasion or manipulation. However, reflecting on ethical debates regarding digital nudging in academia, we find that current conceptualizations are scare. This is where on the basis of existing literature, we provide a conceptualization of ethical guidelines for the design of digital nudges, and thereby aim to ensure the applicability of nudging mechanisms in virtual environments.*

## 1. Introduction

In recent years, "digital nudging" has become an important research focus in the information systems (IS) community (see e.g. [1-4]; [54-61]). Thaler and Sunstein first conceptualized the idea of nudging as a form of overt and predictable behavior change in their work [5]. They state that nudging methods, e.g. encouraging prosocial behavior, can be surprisingly effective, while being libertarian, meaning that they leave people emancipated in their freedom of choice by not excluding any possible choice, nor introducing economic incentives to extrinsically alter behavior [5]. Critics argue that nudging is manipulative in that it undermines people's autonomy [6], by influencing the decision-making process and taking advantage of predictable irrational reactions that result from heuristics, biases and psychological mechanisms ([7]; [8-9]). Individuals are susceptible to be influenced in their decision-making processes as they fall back on heuristics, biases, and psychological mechanisms [2]. Digital nudging is implemented in virtual environments to e.g. simplify the information and option overflow. However, whenever IS design aims to influence human behavior, IS scholars should also address ethical concerns ([10-12]). Against this background, the ethical legitimization of nudging is controversially discussed in literature.

It is important to discuss and define nudging in distinction to other regulatory mechanisms such as manipulation (infringement of autonomy), to better apprehend the ethical debate on nudging. Due to the difficulty to distinguish both concepts and the premise that general ethical abstractions often lead to insufficient and perplexing results [13], we hereby aim to concretize the ethical debate on nudging. More specifically, with the increasing importance of IS at present, and the increasing use of e.g. artificial intelligence, it is important to extend the debate on ethical legitimizations of nudges to the digital context.

Yet, according to Renaud and Zimmermann [14], despite the growing importance of ethical discussions in IS, research on ethics is only slowly emerging. *With this work, we aim to contribute by providing ready to use ethical guidelines, which transfer knowledge derived from definitory and ethical discussions in (offline) nudging to the online context, in order to support designers of digital nudges*. Thereby, we align with the differentiated view on ethics and morality as suggested by Stahl [15] and further discussed by [16]. In literature, ethics and morality are often used interchangeably [17-18] and are jointly described as "customs, good practices or expected behavior" ([19], p. 145). However, the differentiation between morality, understood as rules or norms within a society and ethics, conceptualized as the justification of these morals [15], provides a better generalizability. Namely, that nudge designers with different understandings of morality may still follow the same universal ethical frameworks and therefore collectively find applicability of our guidelines [19]. Moreover, as part of our contribution, the guidelines may also provide a standard upon which designers can be held accountable for.

The paper is structured as follows: in section 2, we will provide the background and basic understandings of the nudging concept by emphasizing on ethical

HICSS

aspects of nudging in regard to libertarian paternalism and manipulation. Thereafter, in section 3, we will present our research design, before we work towards ethical guidelines in digital contexts, which are derived from existing literature in section 4. In section 5, we will present our discussion, followed by a summary and outlook to future research in section 6.

## 2. Theoretical background

### 2.1 Bounded rationality and digital nudging

A short introduction into behavioral science, and existing cognitive processes guiding human decision-making provide a basic understanding of ethical discourses in nudging. Bounded rationality and the underlying idea of decision-making heuristics is based on the assumptions that humans dispose over two interconnecting cognitive systems: the unreflective (system 1) and the reflective (system 2) [5; 20]. Humans most regularly use unreflective (automatic) thinking. It is fast, effortless, relies on cognitive heuristics, and biases [20-21], and enables individuals to process multiple thoughts simultaneously. The reflective system, in contrast, is slow, and effortful as it comprises conscious and deliberate thinking [20-21]. Here, individuals process information sequentially, allowing them to critically reflect and analyze the information. Such 'bounded rationality' [20], is the underlying basis for the concept of nudging.

In this context, a nudge can be defined as "(...) *any aspect of the choice architecture that alters people's behavior in a predictable way without forbidding any options or significantly changing their economic incentives*" ([5], p. 6). It is always intended to either alter a person's unreflective thinking and consequent automatic behavior, or trigger the decision-making processes in the reflective system to guide conscious decisions, therefore altering choice [5]. Transferred to the virtual environment, digital nudging emphasizes on the medium and digital instruments through which nudges can influence choice architectures. It can be defined as the "use of user-interface design elements to guide people's behavior in digital choice environments" [12, p. 1] or a "*a subtle form of using design, information and interaction elements to guide user behavior in digital environments, without restricting the individual's freedom of choice*" [1, p. 3]. The tools and techniques used in digital nudging affect the same psychological mechanisms (i.e. biases, heuristics) as in the offline context. Since digital interfaces are always created by a choice architect and do not appear randomly, it is crucial to understand the design of digital (user) interfaces as the environment which influences decisions [2].

Zhang and Xu [26] found that transparency and the attitude towards privacy settings can be improved, showing the effectiveness of digital privacy nudges in mobile apps. Similarly, Weinmann et al. [12] acknowledged that opt-in or opt-out default settings have a big influence on online user behavior. Default-setting research further reveals that nudges can (1) positively influence environmentally sound decision-making during [27] or (2) decrease (online) procrastination behavior in students [28]. More recent studies on online privacy assessed that the combination of different claims and supporting arguments is most promising to nudge an increase in privacy concerns [2]. Meanwhile, providing information about the incompatibility of goods during the purchasing process improves the online shopping behavior [29], while digital nudging may also influence users to reflect on their own sharing behavior and potentially related privacy concerns [30].

To this day, digital nudging receives less attention in enterprise settings, though it has great potential in working environments [32]. Kretzer and Maedche [3] pre-defined so-called "social nudges" which targeted a variety of different effects such as cohesion, business function, geographical distances, and hierarchy. In an experimental setting, they were then able to show a positive effect of their social nudges on the decision-making of users (e.g. enterprise report recommendations) [3].

### 2.2 Regulatory mechanisms and manipulation

The academic discourse on regulatory mechanisms involves (1) architecture (or code), as an alteration of the physical or digital environment to make alternative decisions less appealing, (2) libertarian paternalism, as a mechanism to positively influence rational decision-making, and (3) mandated disclosure (or notice) entailing the distribution of facts to influence informed decision-making [33]. While code was the first concept to emerge [34], it generally entails a manipulative character as choice architectures are influenced through sanctions [34] or other forms of governmental control to influence citizen behavior such as speed bumps to regulate driving speed [33]. Thereafter, nudging emerged and addressed decision-making based on the bounded rationality of humans as a way of aiding the choice selection of preferential options within existing choice architectures, rather than constructing choice architectures in a mandated and strictly paternalistic way [5]. However, as Thaler and Sunstein acknowledge the (soft) paternalistic character of nudges, they emphasize the libertarian aspect of nudging (e.g. the option to choose differently as nudged by authorities) as the distinguishing factor between nudging and manipulation [33].

A commonly used definition describes manipulation as "*intentionally causing or encouraging people to make the decisions one wants them to make, by actively promoting their (decision-)making (…) in ways that rational persons would not want to make their decisions*" ([35], p. 33). From our point of view, this definition likely applies to the idea of code. Moreover, nudging critics argue that nudging has a manipulative character [36]. Despite the efforts of Thaler and Sunstein [5] to legitimize nudging by distinguishing it from code, there is still continuous criticism regarding a clear understanding of the economic incentives or the lack of defined circumstances for the removal or addition of options [36]. In line with the criticism, the nudging concept does not provide a clarification nor do legitimizing conditions exist to differentiate nudging from manipulation. In the effort of closing this gap, we will review existing literature on more detailed approaches to define nudging and distinguish nudges from manipulation, before building our ethical guidelines on the basis thereof.

Hansen [37] draws the conclusion that there are two possibilities for the definition of a nudge. One being that nudges pursue paternalistic motives, which simplifies their ethical legitimization as the motive of the nudger has to be based on the interests of the nudgee [37]. The other conceptualizing nudging from a technical perspective as the effort to influence human judgement, choice or behavior in predictable ways where rational decision-making in line with one's own preferential choice set falls short to bounded rationality (e.g. due to cognitive boundaries, biases, routines) [37]. In consequence, and contrary to Thaler and Sunstein [5], here nudges (1) exclude relevant choice options, (2) grant extrinsic incentives, and (3) provide rational information and arguments [37]. In this definition, nudges are not necessarily paternalistic, therefore not always align with the preferences of the nudgee. Instead, this definition is more suitable for practical use because it includes both, nudges that have paternalistic motives as well as third-party interests. This is where nudging reaches its conceptual limitations because it is originally not laid out to include non-libertarian phenomena [38], yet, in reality many nudges have long been used for such purposes.

According to Grüne-Yanoff [39], it is impossible to consider all individual nudgee preferences for the construction of a choice architect. Nudging and especially digital nudging, always contain some prejudice towards certain choices. 'Subconscious' nudges often reach the biggest effect [7], which particularly decreases the incentive for designers of digital nudges to aim at transparent implementations. Moreover, the implementation of non-transparent nudges bares one of the most central ethical concerns and reservations against nudging as transparency is one of the most important legitimizing conditions for the distinction between nudging and manipulation [5].

In regard to the ethical debate on nudging, research has provided a few different perspectives. On the basis of her conceptual paper on possible aspects counteracting the side-effects (e.g. distrust) or autonomy concerns of nudging, Clavien [65] argues that shared preference justifications (all affected nudgees agree with the goals pursued by the nudger) are the most adequate to ensure ethical legitimization of nudges. Meanwhile, Blumenthal-Darby and Burroughs [66] focused their work on the identification of ethically relevant dimensions in form of a list of considerations that must be acknowledged when choice architectures are being influenced.

In his analyses of case studies, Engelen [64] establishes a so-called list of criteria and their suggested contextual importance, which health practitioners should consider when nudging their patients. However, in a response to Engelen [64], Fowler and Roberts [67] criticize his work and suggest broadening its perspective to include a wider variety of cases for analyses, on top of the perspectives of those affected by the nudges as well as an assessment of the effectiveness of the implemented nudges. In light of the existing work on digital nudging and ethics it becomes apparent that though there have been some efforts to incorporate ethical standards into nudging, to this day two aspects are still missing: (1) the evaluation of ethical standards in digital nudging and (2) the introduction of ready to use guidelines for the actual design of nudges. This is also supported by Lembcke et al. [68] who point out, that an "*integration of practical ethical guidelines into* (an) *existing framework*" is yet to be done" (p. 13). We agree that the existing body of knowledge is still lacking a set of guidelines that can be utilized by designers in order to allow theory to actually inform practice. That is why our work set out to provide both aspects, thereby closing this gap.

## 3. Method

On the basis of the provided definitions and first aspects to distinguish nudging from manipulation, we now introduce our literature analysis. We analyzed literature on regulatory mechanisms, nudging, and persuasion. We built our methodological approach based on the framework by vom Brocke et al. [40] and aligned our literature review with the categorization by Cooper [41], to assure a justified scope in line with the goal and target group (Fig. 1). The goal of our literature analysis is a thorough representation of the ethical debate on nudging. Moreover, our goal was to review other potential concepts to provide a holistic foundation for the derivation of ethical guidelines. The selection process was not entirely neutral as we pre-selected relevant publications, which fit our goal and aligned with the IS context. With our work we aim to address scholars in nudging related topics and designers or practitioners implementing nudges in virtual environments.

**Table 1**. Visualization and categorization approach (Cooper, [41]; vom Brocke et al., [40]).

| | Character-istics | Categories | | | |
|---|---|---|---|---|---|
| 1 | focus | research outcomes | research methods | theories | appli-cations |
| 2 | goal | integration | criticism | | central issues |
| 3 | organi-zation | historical | conceptual | | method-logical |
| 4 | pers-pective | neutral representation | | espousal of position | |
| 5 | au-dience | specialized scholars | general scholars | practition ers/politician s | general public |
| 6 | cov-erage | exhaus-tive | exhaustive and selective | represent ative | central/pivotal |

Forward and backward searches on all identified research articles allowed us to include relevant literature e.g. on regulatory mechanisms. Hereby we took a particular interest in publications discussing ethics and conceptual research methodologies. We derived our ethical guidelines iteratively by comparing relevant concepts and frameworks (e.g. [68]).

The process of deriving the ethical guidelines for the design of nudges is twofold. We first used a categorization scheme based on the framework as introduced by Hansen and Jespersen [42], distinguishing between transparent and non-transparent nudges in combination with system 1, unreflective and system 2, reflective thinking. In a second step, we closely regarded the identified literature to derive our guidelines in a stepwise approach. Altogether, we used a descriptive research methodology by casting a light on a current issue and gradually building on already existing work. Accordingly, the ethical guidelines we introduce in the following were developed on the basis of past work and the idea, that an ethical approach to the design of nudges should consider transparency and the cognitive thinking processes of the target audience as the two starting points of consideration.

# 4. Results: ethical guidelines for the design of digital nudges

Our guidelines assist and provide a ready to use check-list, to guide researchers and practitioner in the design process of nudges. In line with this goal we provide a universal framework that applies to all digital nudges, and most importantly for all designers independent of their occupational or epistemological background.

## 4.1 Transparency as a legitimizing condition for nudges

While it seems logical in theory that individuals always have the freedom to choose differently than what they are being nudged towards [5], a broad variety of nudges are not transparent and therefore non-paternalistic. Consequently, nudging gives individuals the theoretical option to choose differently, but practically their behavior is changed in a way that does not provide options to vary choices.

Given that nudges always have to be transparent [5], it is crucial to regard transparency in more detail. Namely, (digital) nudges can be distinguished as transparent and non-transparent [43]. It is the non-transparent types of nudges that fall under the manipulation objection as they work subliminally and are usually not striking to the nudgee [46]. Considering the unreflective (system 1) and reflective (system 2) cognitive systems, we recognize that there are different forms of influence between these two types of nudges.

Since system 1 nudges influence (e.g. simplify) choices, they can infringe an individual's autonomy, if undetected by the nudgee [23]. System 2 nudges alter a nudgee's reflective thinking (e.g. habits) [23], possibly influencing a shift in behaviour (e.g. exclude choice). However, non-transparent system 1 nudges can only be legitimate, if preceded by some type of information or clarification about the nudge. This precondition cannot apply to non-transparent system 2 nudges, as the nudger intentionally excludes possible choices, leading to an intentional and unethical manipulation of the contextual environment and consequent behaviour. This concludes that non-transparent nudges have to be used in a disclosed manner and the nudgees have to be able to intervene and resist the triggered behavioral change at all times.

Wilkinson [47] supports this distinguished view of non-transparent system 1 and 2 nudges by stating that nudges, which neither qualify as a form of intentional influence on choice (unreflective system 1) nor inflict on or pervert an individual's decision-making process and consequent behavior (reflective system 2), can be considered manipulation. Therefore, under the given considerations, which we will also further discuss as "legitimizing conditions", ethical nudges can either be transparent or non-transparent (see e.g. Table 3).

## 4.2 Easy-resistibility as a legitimizing condition for (non-)transparent nudges

In line with Thaler and Sunstein [5], the original choice sets of a nudged person are limitless and should not be inhibited by external factors. Accordingly, Saghai [48] suggests, that if choice architects are used to shape choices, users must also be given the option to resist. Thaler and Sunstein [5] agree by stating that nudges should be "easy and cheap to avoid" (p. 9).

According to Saghai [48], easy resistibility is provided, when a nudgee has the ability and chance to become aware of how the nudge is steering the individual towards a certain behavior; when a nudgee has the ability to oppose the triggered behavior or choice; or when the influence of the nudge is not undermining the nudgee's attention-bringing or inhibitory capacities. We conclude that non-transparent system 1 nudges can only be legitimate, if easy resistibility is ensured. Meanwhile, non-transparent system 2 nudges are excluded from this legitimization as a nudgee has no ability to oppose influence taking he or she is unaware of.

## 4.3 Non-controllability as a legitimizing condition for (non-)transparent nudges

Moreover, Saghai [48] also states that substantially non-controlling nudges, shaping the environment through intentional choice architectures may be libertarian. This means, that the nudger should not take any measures to violate the autonomy of the nudgee, such e.g. incentives or coercion.

The substantial non-controlling condition therefore protects the nudgee from influences that change the individual's perspective in a subliminal or even manipulative way. However, as it can be argued that resistibility is a subjective criterion, following the argument by Faden and Beauchamp [49], Saghai [48] suggests that objective views of resistibility align with the perception of the average person. We summarize that to distinguish digital nudging and (non-libertarian) manipulation, we first acknowledge prejudices as a constituent of all decision-making environments, especially in the digital world.

Further, we acknowledge, that if nudgers construct choice architectures they have to do so transparently,

include a possibility of resistance, and avoid extrinsic incentives. This is also expressed by the term "easy resistibility", which is often used in the nudging context. One may hence speak of manipulation when a non-transparent nudge is implemented without any preceding information, and is further perceived as an unwanted, disfavored alteration of decision-making processes in digital environments. According to our provided arguments and conditions to distinguish paternalism from manipulation, Table 2 provides an overview and categorization scheme of the essential conditions for legitimate nudges as well as legitimizing conditions for non-transparent nudges that we derived from the literature. Further, Table 2 also resembles the basic frame, on the basis of which we derived our ethical guidelines in Table 3. Nudge designers need to be mindful of the public welfare or normative incentive applying to their target group (Step 1) [69], and use this knowledge as the basis on which nudging goals (Step 2) and design principles (Step 3) can be derived, before their evaluation (Step 4).

**Table 2. Summary of ethically legitimizing conditions of nudges**

| Essential conditions | Legitimizing Conditions | Authors |
|---|---|---|
| Libertarian Paternalism | Transparency | [43]; [5]; [37] |
| | System 1: easy resistibility, non-controlling | [42]; [7]; [46]; [50] [48] |
| | System 2: easy resistibility, non-controlling | [48] |
| | Non-transparency | |
| | System 1: disclosure, consent, easy resistibility, non-controlling | [50]; [42]; [47] |
| | System 2: nudges are manipulative and beyond legitimization | [51]; [47]; [42] |

Hence, we distinguish nudges that affect the reflective thinking from manipulative nudges that affect the unreflective thinking, if they are non-transparent or according to [46] remain undetected by the nudgee. In regards to non-transparent nudges we are now able to further derive legitimizing conditions of ethical nudges. Based on existing literature regarding the basic characteristics of (digital) nudging as well as the described legitimizing conditions, Table 3 provides ethical guidelines in form of a checklist that has to be satisfied in order to assure the design of ethical digital nudges.

**Table 3. Ethical Guidelines for the Construction of Digital Nudges**

| Ethical Guidelines for Digital Nudging: A stepwise approach | | | |
|---|---|---|---|
| **Step 1:** Understand the intentions of potential users and their cognitive heuristics and biases | ☐ the target group has been thoroughly identified<br><br>☐ the preferential choice set of your target group has been identified | | |
| **Step 2:** Derive the goals of digital nudging | ☐ the goals are in alignment with the users' preferential choice set and/ or stem from good intentions. They benefit and do not harm the user.<br>☐ the potential impact is predictable | | |
| **Step 3:** Design and implement the nudge | | | |
| *System 1 (Unreflective Thinking) fast, unconscious decision making & parallel/ convergent thinking* | | | |
| **Transparent** | ☐ the choice architecture is presented in the most simplified way | **Non-transparent** | ☐ justification for the need for non-transparency is given |
| Easy resistibility | ☐ the nudge is easy and cheap to avoid. There are no costs to avoiding the nudge. | Disclosure | ☐ simplified information about the nudge is provided |
| | | Consent | ☐ consent forms are provided requiring users to thoroughly read & opt-in to the terms and conditions to ensure informed consent, or informational nudges signaling preselected default-settings are provided, in cases were informed consent was already established |
| Non-controlling | ☐ no incentive/ coercion was introduced to influence choice | Easy resistibility | ☐ the nudge is easy and cheap to avoid. There are no costs to avoiding the nudge. |
| | | Non-controlling | ☐ no incentive/ coercion was introduced to influence choice. |
| *System 2 (Reflective Thinking) slow, conscious & sequential/ critical thinking* | | | |
| **Transparent** | ☐ the choice architecture is presented in a simple and comprehensive way | **Non-transparent** | Designing non-transparent digital system 2 nudges (reflective) is considered manipulative. Therefore, the nudge should now be re-considered. |
| Easy resistibility | ☐ the nudge is easy and cheap to avoid. There are no costs to avoiding the nudge. | | |
| Non-controlling | ☐ no incentive/ coercion was introduced to influence choice. | | |
| **Step 4:** Evaluation of the digital nudge and iteration | ☐ the nudge is consistent with the original goal and useful to influence the target behavior<br>☐ there are no unintended negative consequences (e.g. malicious intent, monetary disadvantages) for the target group. | | |

# 5. Discussion

As supported by research of [1] and [2], the nudger has to understand the intentions of potential users and their heuristics as well as biases in the first step. This includes acknowledging that the targeted nudgees may experience both, unreflective and reflective thinking during their interaction with the nudge and consequent decision-making process. In the consecutive second step the nudger can now derive the goal a digital nudge aims to achieve. On the basis thereof the nudger is able to design and implement the nudge. Hereby, the nudger needs to be aware of the different cognitive systems his target group fluently switches in and out of when interacting with the digital interface.

When implementing **transparent digital system 1 nudges** (unreflective), automated behavior is being influenced, making these types of nudges almost impossible to avoid. Although they appear to be detectable to the nudgee, they work manipulative in a technical (not psychological) way. As they are overt and identifiable, these nudges are ethically justifiable. Here, the choice architect is held responsible for the nudging effects because an individual's automatic thinking guides the user behavior. If an individual is nudged against personal preferences, attention bringing capacities can be activated and deliberate thinking allows the individual to resist the nudge [48]. If the goal of a choice architect is to e.g. increase the security settings during online purchasing, respective security settings could be nudged in a simplified and accessible way (e.g. reminders to update settings or increased visibility of security options) that the nudgee may automatically feel encouraged to update the privacy settings.

When designing **non-transparent digital system 1 nudges** (unreflective) manipulation risks are high, as they are almost undetectable and influence behavior subliminally. Thus, it is difficult for the nudgee to reconstruct the ends and means of the nudge or perceive all possible options. Hence, choice architects should always aim to be in line with user interests. However, as individual user interests may be difficult to reconstruct for designers, another option is to reveal the intentions behind the nudge or request the user's consent. This would require the nudger to implement at least one of the following legitimizing conditions: disclosure of information or asking user consent as a prerequisite to continue using the website. The mentioned legitimizing conditions, also known as informed consent [14] are however not always prompted, especially given the vastness by which users move across digital landscapes, oftentimes holding multiple registrations with different platform providers. In such cases we acknowledge that a ready-to-use ethical guideline needs to be aware of the discrepancy between ethical nudging and providing the least amount of disruption during a user experience. Therefore, instead of suggesting to request informed consent every time a user logs onto a platform, we suggest that priming or signaling nudges could be implemented to raise the user's awareness. This would qualify for an information box signaling the users that he was nudged according to the preselected default-settings that were agreed upon during the initial registration process on the current platform for instance.

Designing and implementing **transparent digital system 2 nudges** (reflective), entails influencing choice in a transparent way, by emphasizing choices consistent with the nudgee's preferences [42]. This system of nudges activates reflective thinking, making it easy to understand the means and ends of the implemented nudge and giving the nudgee the actual possibility to make choices without perverting his or her decision-making process. These nudges are truly libertarian as they influence individuals without manipulation or infringement of autonomy. This is in line with Tocchetto [63] who argues that only reflective (and transparent) nudges are morally harmless. Here, the nudger would e.g. use timely information nudges as triggers reminding the user of the insufficient security settings or provide additional information during the user journey on the website to encourage the nudgee to reflect password or security choices.

Designing **non-transparent digital system 2 nudges** (reflective) on the other hand is considered non-paternalistic and manipulative. Although these nudges use tools that activate reflective thinking, they do not allow the nudgee to reconstruct the means and ends of the implemented nudge. Choices are manipulated, leading to behavioral change without the opportunity to choose differently. Here, nudgees would be forced to update or revisit security settings before e.g. logging out. Even though this could potentially increase the overall security settings it would manipulate user behavior and infringe freedom of choice.

Finally, the fourth and last step in the design or implementation of ethical nudges is to evaluate the nudge and, if necessary, interactively adjust aspects according to the principles we have provided. This step will be subject to further research. We used a heterogenous body of literature as part of our methodology to derive the ethical guidelines.

Therefore, as a final step, we added the evaluation and possible iteration of the digital nudge. By doing so we adapted known evaluation processes as part of IS research affiliated Design Science Research methodologies (see e.g. [52]), which provides an important implication for nudging research and its constant effort of putting nudging theory into practice. We view the evaluation phase of the ethical guidelines as a central final step to assure functional purpose and target analysis [52] and provide a quality standard as part of our key practical contribution to establish ethical nudging design guidelines. Namely, the evaluation is a reassessment if the results match the expectations (steps 1-3) [53]. After we derived and iteratively improved the ethical guidelines for nudging, we aim to test and evaluate our guidelines in the future (e.g. by designers of digital nudges) to provide legitimization for the use of digital nudges in IS specific contexts.

## 6. Conclusion

In this paper, we critically reflected the current status quo in the ethical discussion on digital nudging. We expanded on existing research by drawing between libertarian nudges, non-libertarian nudges and manipulation. The latter only being present in cases where non-transparent nudges are implemented in the attempt to alter people's decision-making processes and consequent behaviour in a way they would not have chosen by themselves, thus violating their autonomy and freedom of choice.

We conclude, that designers of a choice architecture have to be able to defend the used measures, thereby self-checking if their nudge withstands public scrutiny. Our literature-based conceptualization of ethical guidelines very well reflects the status quo of the existing academic discourse. However, we acknowledge that both, providing different case scenarios of the application of our guidelines as well as possibly mapping these to the structure of a nudging design method, therefore allowing us to derive an actual research question, would highlight the applicability of our guidelines and further improve the contribution to practice. This is why we will consider both aspects in future research. As a next step, the guidelines need empirical validation and discussion with nudgers and nudgees. We hence suggest implementing and testing our guidelines in a consecutive research project to expand on the contribution for practitioners, validate our guidelines, and provide a concretized call to action for their enforcent.

## 7. Acknowledgements

We thank Nico Kastunowicz for his support.